\documentclass[preprint,amsmath,amssymb,showpacs,groupedaddress,byrevtex]{revtex4}
\usepackage{graphicx}
\usepackage{dcolumn}
\usepackage{bm}
\usepackage{epsfig}
\usepackage{amsmath}

\begin{document}
\draft

\def\centereps#1#2#3{\vskip#2\relax\centerline{\hbox to#1{\special
  {eps:#3 x=#1, y=#2}\hfil}}}

\newcommand{\bq}{\begin{subequations}}
\newcommand{\eq}{\end{subequations}}
\newcommand{\bqq}{\begin{subeqnarray}}
\newcommand{\eqq}{\end{subeqnarray}}
\newcommand{\beq}{\begin{eqnarray}}
\newcommand{\eeq}{\end{eqnarray}}
\newcommand{\beqq}{\begin{eqnarray*}}
\newcommand{\eeqq}{\end{eqnarray*}}
\newcommand{\be}{\begin{equation}\displaystyle}
\newcommand{\ee}{\end{equation}}

\newcommand{\st}[1]{{\mbox{${\mbox{\scriptsize #1}}$}}}  
\newcommand{\uv}[1]{\mbox{$\widehat{\mbox{\boldmath $#1$}}$}}   
\newcommand{\dy}[1]{\mbox{\boldmath $\overline{#1}$}}   
\newcommand{\nab}{\mbox{\boldmath $\nabla$}}            
\newcommand{\gleq}{\mbox{$\begin{array}{c}>\vspace{-2mm}\\<\\ \end{array}$}}
\newcommand{\CS}{\mbox{$\begin{array}{c}\cos \\ \sin \\ \end{array}$}}
\newcommand{\SC}{\mbox{$\begin{array}{c}\sin \\ \cos \\ \end{array}$}}
\newcommand{\mt}{\mbox{$\hspace{-1.5ex}$}}              
\newcommand{\Eo}[1]{{\mbox{$^{e}_{o}{\scriptstyle #1}$}}} 
\newcommand{\Oe}[1]{{\mbox{$^{o}_{e}{\scriptstyle #1}$}}} 

\title{Achieving Invisibility of Homogeneous Radially Anisotropic Cylinders by Effective Medium Theory}
\date {\today}
\author {Y.X. Ni$^1$, L. Gao$^1$\footnote{E-mail:
leigao@suda.edu.cn}, and C.W. Qiu$^2$\footnote{E-mail:
cwq@mit.edu}}
\address {$^1$ Jiangsu Key Laboratory of Thin Films, Department of Physics, Soochow University, Suzhou 215006, China \\
$^2$ Research Laboratory of Electronics, Massachusetts Institute of Technology, 77 Massachusetts Avenue, Cambridge MA 02139, US}




\begin{abstract}
In this paper, we establish the full-wave electromagnetic
scattering theory to study the electromagnetic scattering from
infinitely long cylinders with radially anisotropic coatings. We
show that the total effective scattering width can be dramatically
reduced by the suitable adjustment of the dielectric anisotropy of
the shell, while it is not the case for tuning the dielectric
anisotropy of the core. Furthermore, we could make the cylindrical
objects invisible when both dielectric and magnetic anisotropies
are adjusted. In the long wavelength limit, we develop effective
medium theory to derive the effective isotropic permittivity and
permeability for the anisotropic coated cylinders, and the
invisibility radius ratio derived from the full-wave theory for
small coated cylinders can
be well described within the effective medium theory.
\end{abstract}

\maketitle
\newpage

\section{Introduction}
In recent years, designing optical and electromagnetic (EM) invisibility
cloaking devices has attracted great attention from both physics and engineering societies. Scientists have developed coordinate transformation method \cite{Pendry} and conformal mapping method \cite{Leonhardt}, which can
protect the cloaked object of arbitrary shape from electromagnetic
radiation. This idea has been verified by full-wave simulation
of the cylindrical cloaking structure based on both finite element
method (FEM)~\cite{Cummer} and finite-difference time-domain
method (FDTD)~\cite{Zhao}. The experimental demonstration at
microwave frequencies~\cite{Schurig} and even at optical
frequencies~\cite{Cai1} for non-magnetic material has been
realized lately. They have investigated
the transformation based cloaks possessing spherical/cylindrical geometries
with rotational symmetry~\cite{Pendry,Schurig,Ruan,Chen1,Jiang} and
geometries with reduced symmetries, such as eccentric elliptic
cylinders~\cite{Kwon}, square cloaks~\cite{Rahm}, and arbitrary shapes \cite{numerical1}. In addition to the method of spatial transformation, Alu {\itshape et. al.} have proposed to use isotropic plasmonic coatings to render objects invisible~\cite{Alu2} based on the dipolar cancellation, and
   proposed a parallel-plate metamaterial cloak~\cite{Silveirinha} that significantly reduces the total scattering
   cross section of a given two-dimensional dielectric obstacle in some frequency band.

Due to the spatial compression, the required parameters should be anisotropic and dependent on the position \cite{Pendry}. To alleviate these contraints, EM invisibility cloaks have been realized by isotropic coatings based on the effective medium theory \cite{Gao,Qiu_PRE2009}. However, it still requires sufficient layers of alternating isotropic media in order to maintain the validity of that effective medium theory. Although recent development in metamaterial may allow the control of material parameters, such strict requirements for the parameters of a perfect cloak are still too difficult to realize in practice. As a way to reduce the number of required constitutive
parameters to realize a manufacturable cloak, the incidence can be decomposed into transverse electric (TE) and transverse magnetic (TM) components so that 2D cloaks with simplified parameters \cite{Yan1,Luo} can be used. Unfortunately, the simplified
cylindrical cloak will cause large reflection at the cloak's outer boundary due to the impedance mismatch. From the discussion above, one can see that: (1) ideal cylindrical cloaks require 6 parameters, some of which are infinite values; (2) simplified cylindrical cloaks requires 3 parameters but still some value tends to the infinity at the inner boundary; (3) both ideal and simplified cylindrical cloaks need certain parameters to be inhomogeneous i.e., a function of position. This motivates our current work on achieving cylindrical cloaks, which is based on only one single coating by a homogeneous and radially anisotropic medium. The invisibility condition can be established by the effective medium theory. It is obvious that the contraints in material fabrication are greatly alleviated.

A lot of theoretical and numerical approaches have been developed to deal with the scattering problem of homogeneous anisotropic or even gyrotropic cylinders, e.g., integrodifferential equation \cite{Uslenghi,Uslenghi_2}, volumetric integral equation method \cite{ZNChen_1}, combined field surface integral equation method \cite{ZNChen_2}, finite difference method with measured equation of invariance (FD-MEI) \cite{ZNChen_1}, and dyadic Green's functions \cite{cttai:1994,nkuzunoglu:1995,Qiu_PRB_2006}. However, it should be noted that all these anisotropic tensors in those mentioned papers are defined in Cartesian coordinates, which is different from the radially anisotropy defined in the cylindrical coordinates as discussed in this paper. Although the latter can be mapped into Cartesian anisotropy, then the anisotropic cylinder will be inhomogeneous and angle-dependent.

Following our discussed motivation, in order to investigate the EM invisibility of radially anisotropic coated cylinders, we develop the compact scattering theorem for a coated cylinder with homogeneous radial anisotropy, by extending the idea of embedding radial anisotropy in the orders of Bessel/Hankel functions in spherical case \cite{Qiu_TAP_2007}. We focus our analysis on the invisibility characteristics of a homogeneous anisotropic coated cylinder. EM field components and scattering width of such coated cylinders are formulated. We discuss the roles of anisotropic parameters as well as the core-shell ratio on the reduction of the total scattering section. From the scattering algorithm, the effective permeability and permittivity of the core-shell system are also established so that the required condition for invisibility performance can be exactly determined. The numerical results are given.


\section{Full-wave electromagnetic scattering theory}

Let us consider the electromagnetic scattering from a
radially-anisotropic coated cylinder of infinite length (see
Fig.~1).

\begin{figure}[htbp]
\begin{center}
\includegraphics[width=7cm]{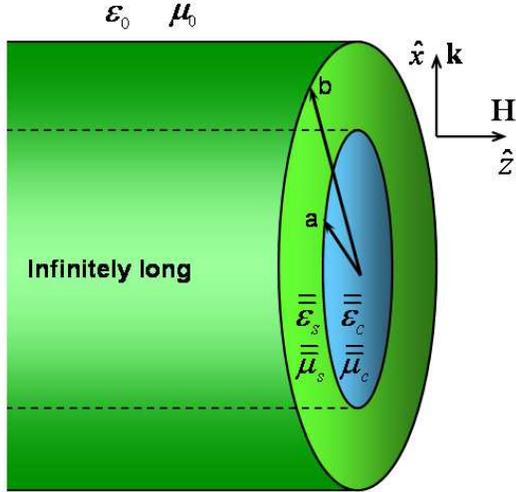}
\caption{Geometry of the scattering of a plane wave by a coated
cylinder with permittivity and permeability of radially anisotropic tensors. The incident
wave propagates along x-axis, and the magnetic field is along z-axis.}\label{Geo}
\end{center}
\end{figure}

For the simplicity, we assume a normally incident plane wave with
the magnetic field along the z direction (i.e., the transverse
magnetic (TM) wave). We assume that the
coated cylinder is composed of  the core with the radius $a$ and
the shell with the radius $b$, and it is surrounded by free space
$(\varepsilon_{0},\mu_{0})$. As for radial anisotropy, we indicate
that the permeability and the permittivity tensors for the core
and the shell can be  expressed in cylindrical coordinates
$(\hat{r},\hat{\theta},\hat{z})$, \bq \beq \overline{\overline
\varepsilon } _p  \mt& =&\mt \left( {\begin{array}{*{20}c}
   {\varepsilon _{pr} } & 0 & 0  \\
   0 & {\varepsilon _{p\theta } } & 0  \\
   0 & 0 & {\varepsilon _{pz} }  \\
\end{array}} \right) \\
  \overline{\overline \mu } _p  \mt& =&\mt \left( {\begin{array}{*{20}c}
   {\mu _{pr} } & 0 & 0  \\
   0 & {\mu _{p\theta } } & 0  \\
   0 & 0 & {\mu _{pz} }  \\
\end{array}} \right),
\eeq
\eq
where $ \overline{\overline \varepsilon } _p$ (or
$\overline{\overline \mu } _p$) represents the permittivity (or
permeability) tensor of the core by $p=c$ and/or the shell by
$p=s$.

In Fig.~1, the incident magnetic field with unit
amplitude can be written as $H_z  = e^{ik_0 x-i\omega t}$, where
$k_0 = \omega \sqrt {\varepsilon _0 \mu _0 }$. In what follows,
$e^{- i\omega t}$  will be omitted.  In this connection, Maxwell
equations in the core and shell are written as,
\bq\beq
\nabla  \times{\bf {H}} \mt& =&\mt  - i\omega \overline{\overline
\varepsilon } \cdot{\bf {E}}\\
 \nabla
\times{\bf {E}}  \mt&=&\mt  i\omega \overline{\overline \mu } \cdot{\bf
{H}}.
\eeq\eq

In cylindrical coordinates, Eq. (2) can be decomposed into
\bq\beq
i\omega \mu _{pr} H_r \mt& =&\mt \frac{1}{r}\frac{{\partial E_z
}}{{\partial \theta }} - \frac{{\partial E_\theta  }}{{\partial
z}}\\
 i\omega \mu _{p\theta } H_\theta \mt& =&\mt \frac{{\partial E_r
}}{{\partial z}} - \frac{{\partial E_z }}{{\partial r}}\\
i\omega \mu _{pz} H_z   \mt& =&\mt  \frac{1}{r}\frac{{\partial (rE_\theta
)}}{{\partial r}} - \frac{1}{r}\frac{{\partial E_r }}{{\partial
\theta }}\\
- i\omega \varepsilon _{pz} E_z \mt& =&\mt \frac{1}{r}\frac{{\partial (rH_\theta  )}}{{\partial r}} -
\frac{1}{r}\frac{{\partial H_r }}{{\partial \theta
 }}  \\
 - i\omega \varepsilon _{pr} E_r \mt& =&\mt \frac{1}{r}\frac{{\partial H_z }}{{\partial \theta }} - \frac{{\partial H_\theta  }}{{\partial
 z}}\\
 - i\omega \varepsilon _{p\theta } E_\theta \mt& =&\mt \frac{{\partial H_r }}{{\partial z}} - \frac{{\partial H_z }}{{\partial
 r}}.
\eeq
\eq

After some algebraic manipulations, we can obtain the governing
equation of $H_z$
\begin{eqnarray}
\frac{1}{{\varepsilon _{p\theta } }}\frac{\partial }{{\partial
r}}(r\frac{{\partial H_z }}{{\partial r}}) +
\frac{1}{{r\varepsilon _{pr} }}\frac{{\partial^2 H_z }}{{\partial
\theta^2 }} + \omega ^2 \mu _{pz} rH_z  = 0.
\end{eqnarray}
Inserting the solution in the form of $ H_z = \Psi (r)\Theta (\theta )$
and introducing a seperation term $m^2$, one obtains the ordinary
differential equation for $\Psi (r)$
\begin{eqnarray}
r^2 \frac{{d^2 \Psi (r)}}{{dr^2 }} + r\frac{{d\Psi (r)}}{{dr}} +
\left(\omega ^2 \varepsilon _{p\theta } \mu _{pz} r^2  -
\frac{{m^2 }}{{\varepsilon _{pr} /\varepsilon _{_{p\theta } }
}}\right)\Psi (r) = 0.
\end{eqnarray}
For the part $\Theta (\theta )$, we have the form $e^{im\theta} $.
The general solution to Eq.~(5) is $ AJ_{pm} (k_p r) + BN_{pm}
(k_p r) $, where $J_{pm}$ and $N_{pm}$ respectively denote the
${m}$th-order Bessel and Neumann functions with the argument of $k_p r$, $k_p ^2 = \omega ^2 \varepsilon _{p\theta } \mu _{pz} $, and $ (pm)^2 =\frac{{m^2 }}{{(\varepsilon _{pr} /\varepsilon _{p\theta } )}}$.
Note that if it is an isotropic case (i.e., $\varepsilon_{pr}=\varepsilon_{p\theta}$), $pm$ reduces to be an integer.

In order to match the boundary conditions at cylindrical
interfaces, the incident magnetic field can be expanded as follows,
\begin{eqnarray} H^{inc}_z  = e^{ik_0
x}  = e^{ik_0 r\cos \theta }  = \sum\limits_{m =  - \infty }^{ +
\infty } {i^m J_m (k_0 r)e^{im\theta } }.
\end{eqnarray}

The scattered magnetic field in each region can thus be formulated
\bq\beq
H_z \mt& =&\mt \sum\limits_{m =  - \infty }^{ + \infty } {A_m J_{cm}
(k_c
r)} e^{im\theta } ,\quad r < a, \\
H_z \mt& =&\mt  \sum\limits_{m =  - \infty }^{ + \infty } {i^m [B_m
J_{sm}
(k_s r)}  + C_m N_{sm} (k_s r)]e^{im\theta } ,\quad a < r < b, \\
H_z \mt& =&\mt  \sum\limits_{m =  - \infty }^{ + \infty } {i^m [J_m (k_0
r)}  + D_m H_m ^{(1)} (k_0 r)]e^{im\theta } ,\quad r > b,
\eeq\eq
where $ A_m $, $B_m $, $C_m $, and $D_m$ are the unknown
coefficients to be determined, and $H_m ^{(1)}$ represents the
$m$th-order Hankel function of the first kind.

Applying the boundary conditions of $
E_\theta$ and $ H_z $ being continuous at $r=a$ and $r=b$, we can
derive the scattering coefficient $D_m$ to compute the farfield parttern
\begin{equation}
D_m=\frac{\left|
\begin{array}{cccc}
  J_{cm}(k_c a) & -J_{sm}(k_sa) & -N_{sm}(k_s a) & 0 \\
  \frac{\varepsilon_{s\theta}}{\varepsilon_{c\theta}}k_cJ_{cm}^{'}(k_c a) & -k_sJ_{sm}^{'}(k_s a) & -k_s N_{sm}^{'}(k_s a)& 0 \\
  0  & J_{sm}(k_s b) &  N_{sm}(k_sb) & J_m(k_0 b) \\
  0 & \frac{\varepsilon_0}{\varepsilon_{s\theta}} k_s J_{sm}^{'}(k_s b) & \frac{\varepsilon_0}{\varepsilon_{s\theta}} k_s N_{sm}^{'}(k_s b)
   & k_0 J_m^{'}(k_0 b)\\
\end{array}\right|}
{\left|
\begin{array}{cccc}
 J_{cm}(k_c a) & -J_{sm}(k_sa) & -N_{sm}(k_s a) & 0 \\
  \frac{\varepsilon_{s\theta}}{\varepsilon_{c\theta}}k_cJ_{cm}^{'}(k_c a) & -k_sJ_{sm}^{'}(k_s a) & -k_s N_{sm}^{'}(k_s a)& 0 \\
  0  & J_{sm}(k_s b) &  N_{sm}(k_sb) & -H^{(1)}_m (k_0 b) \\
  0 & \frac{\varepsilon_0}{\varepsilon_{s\theta}} k_s J_{sm}^{'}(k_s b) & \frac{\varepsilon_0}{\varepsilon_{s\theta}} k_s N_{sm}^{'}(k_s b)
   & -k_0 H^{'(1)}_m(k_0 b)\\
\end{array}\right|}
,
\end{equation}
where the prime denotes the
derivative with respect to the argument. Note that
other coefficients can also be solved simultaneously. and hence the
electric and magnetic fields in each region are obtained. The scattering problem for the transverse electric (TE) case, can be solved in a similar way and the corresponding scattering
coefficients can be obtained by the duality of $ \varepsilon \to \mu$ and
$ \mu \to\varepsilon $.

Scattering and extinction efficiencies are expressed through
scattering amplitudes \cite{VAN DE HULST},
\begin{eqnarray} Q_{sca} = \frac{2}{{k_0 b}}\sum\limits_{m
= - \infty }^\infty {\left| {D_m } \right|} ^2 \qquad {\rm and}
\qquad Q_{ext}  = \frac{2}{{k_0 b}}\sum\limits_{m =  - \infty
}^\infty {{\mathop{\rm Re}\nolimits} (D_m } ).
\end{eqnarray}

\section{Effective medium theory in long wavelength limit}

In this section, we present the formulation of our
effective medium theory for the coated cylinders in the
long wavelength limit, i.e., $k_0 b \ll 1 $
and $k_s b \ll 1 $. As a result, the higher-order moments
proportional to $(k_i b)^{2l+1}$ ($i=o, c, s$) are expected to be
negligible and the effective scattering width of the coated
cylinder is dominated by $m=0$ and $m=1$ terms \cite{Wu,Luk'yanchuk}. Thus, we set the conditions for an effective
medium as  $D_0=0$ and $D_1=0$ \cite{Wu}, corresponding to
\begin{eqnarray} \left| {\begin{array}{*{20}c}
   {J_0 (k_c a)} & { - J_0 (k_s a)} & { - N_0 (k_s a)} & 0  \\
   {\frac{{\varepsilon _{s\theta } }}{{\varepsilon _{c\theta } }}k_c J_0 ^\prime  (k_c a)} & { - k_s J_0 ^\prime  (k_s a)} & { - k_s N_0 ^\prime  (k_s a)} & 0  \\
   0 & {J_0 (k_s b)} & {N_0 (k_s b)} & {J_0 (k_0 b)}  \\
   0 & {\frac{{\varepsilon _0 }}{{\varepsilon _{s\theta } }}k_s J_0 ^\prime  (k_s b)} & {\frac{{\varepsilon _0 }}{{\varepsilon _{s\theta } }}k_s N_0 ^\prime  (k_s b)} & {k_0 J_0 ^\prime  (k_0 b)}  \\
\end{array}} \right| = 0,\label{D0}
\end{eqnarray}
and \begin{eqnarray} \left| {\begin{array}{*{20}c}
   {J_{c1} (k_c a)} & { - J_{s1} (k_s a)} & { - N_{s1} (k_s a)} & 0  \\
   {\frac{{\varepsilon _{s\theta } }}{{\varepsilon _{c\theta } }}k_c J_{c1} ^\prime  (k_c a)} &
   { - k_s J_{s1} ^\prime  (k_s a)} & { - k_s N_{s1} ^\prime  (k_s a)} & 0  \\
   0 & {J_{s1} (k_s b)} & {N_{s1} (k_s b)} & {J_1 (k_0 b)}  \\
   0 & {\frac{{\varepsilon _0 }}{{\varepsilon _{s\theta } }}k_s J_{s1} ^\prime  (k_s b)} & {\frac{{\varepsilon _0 }}{{\varepsilon _{s\theta } }}k_s N_{s1} ^\prime  (k_s b)} & {k_0 J_1 ^\prime  (k_0 b)}  \\
\end{array}} \right| = 0,\label{D1}
\end{eqnarray}
where the subscripts $ s1 = \sqrt {\varepsilon
_{s\theta}/\varepsilon_{sr}} $ and $ c1 =  \sqrt {\varepsilon
_{c\theta}/\varepsilon_{cr}} $ . In the limit of  $k_0 b \ll 1 $
and $k_s b \ll 1 $, we can use the following approximations $ J_0
(x) \cong 1$, $N_0 (x) \cong \frac{2}{\pi }\ln (x/2) $, $ J_0
^\prime (x) \cong  - x/2 $, $ N_0 ^\prime  (x) \cong \frac{2}{{\pi
x}} $, $ J_1 (x) \cong \frac{x}{2} $, $J_1 ^\prime (x) \cong
\frac{1}{2} $, $J_\nu ^\prime  (x) \cong \frac{{\nu J_\nu  }}{x}
$, and $ N_\nu ^\prime (x) \cong  - \frac{{\nu N_\nu }}{x}
$ \cite{JDJackson}. Substituting these approximations into Eqs.~(\ref{D0}) and (\ref{D1}) and replacing $\varepsilon_0$, $\mu_0$ by $\varepsilon_{eff}$, $\mu_{eff}$ respectively, one can obtain
\begin{eqnarray}\mu _{eff} = \mu _{sz} (1 - \frac{{a^2 }}{{b^2 }})
+ \frac{{a^2 }}{{b^2 }}\mu _{cz},\label{mu_eff}
\end{eqnarray} and
\begin{eqnarray}
\varepsilon _{eff}  = \frac{{\varepsilon _{s\theta } [(c1 \cdot
\varepsilon _{s\theta }  + s1 \cdot \varepsilon _{c\theta } ) -
(\frac{a}{b})^{2 \cdot s1} (c1 \cdot \varepsilon _{s\theta }  - s1
\cdot \varepsilon _{c\theta } )]}}{{s1[(c1 \cdot \varepsilon
_{s\theta }  + s1 \cdot \varepsilon _{c\theta } ) +
(\frac{a}{b})^{2 \cdot s1} (c1 \cdot \varepsilon _{s\theta }  - s1
\cdot \varepsilon _{c\theta } )]}}.\label{eps_eff}
\end{eqnarray}
 It is evident that in the long wavelength limit, the electric and
magnetic fields are decoupled. For an isotropic case
$\varepsilon_{pr}=\varepsilon_{p\theta}$ ($p=c,~s$), they reduce to
the results in Ref.~\cite{Wu}.  As a consequence, the coated
anisotropic cylinder in the long wavelength limit can be viewed as an
effective homogeneous cylinder, and the scattering efficiency is
expressed in a simpler form as follows,
\begin{eqnarray}
Q_s  = \pi ^5 (\frac{b}{{\lambda _0 }})^3 (\left| {\frac{{\mu _0 -
\mu _{eff} }}{{\mu _0 }}} \right|^2  + 2\left| {\frac{{\varepsilon
_0  - \varepsilon _{eff} }}{{\varepsilon _0  + \varepsilon _{eff}
}}} \right|^2 ).
\end{eqnarray}

Furthermore, the invisibility condition implies that
$\varepsilon_{eff}$ and $\mu_{eff}$ in Eqs.~(\ref{mu_eff}) and
(\ref{eps_eff}) should be identical to the parameters of the host
medium, i.e., $\varepsilon_0$ and $\mu_0$. Thus, the relation
between core-shell ratio and the anisotropic parameters in the
core and the shell can be drawn \beq\label{condition1} \frac{a}{b}
\mt&=&\mt \sqrt {\frac{{\mu _{sz} - \mu _0 }}{{\mu _{sz}  - \mu
_{cz} }}}, ~~~\mbox{purely~magnetic} \eeq \beq\label{condition2}
\frac{a}{b}  \mt&=&\mt \left[\frac{{(c1 \cdot \varepsilon
_{s\theta }  + s1 \cdot \varepsilon _{c\theta } )(\varepsilon
_{s\theta }  - \varepsilon _0  \cdot s1)}}{{(c1 \cdot \varepsilon
_{s\theta }  - s1 \cdot \varepsilon _{c\theta } )(\varepsilon
_{s\theta }  + \varepsilon _0  \cdot
s1)}}\right]^{1/(2s1)},~~~\mbox{purely~nonmagnetic}. \eeq
In fact, the invisibility conditions above possess a
physical constraint, i.e., $0\leq a/b \leq 1$ according to the set-up in Fig.~1, which implies that
given a certain set of anisotropic parameters for the shell and
core, there may be no invisibility no matter how the core-shell
ratio is tuned. It is also worth noting that if the coated
cylinder is isotropic, Eqs.~(\ref{condition1}) and
(\ref{condition2}) respectively become \beq \frac{a}{b} = \sqrt
{\frac{{\mu _s  - \mu _0 }}{{\mu _s  - \mu _c }}}, \eeq \beq
\frac{a}{b} = \left[\frac{{(\varepsilon _s  + \varepsilon _c
)(\varepsilon _s  - \varepsilon _0 )}}{{(\varepsilon _s  -
\varepsilon _c )(\varepsilon _s  + \varepsilon _0
)}}\right]^{1/2}, \eeq which are in accordance with the results in
Ref. \cite{Alu2}.

\section{Numerical Results}
Based on our theoretical results, we provide
numerical calculations of the scattering efficiencies under different radial anisotropies and
physical insights into the invisibility phenomena.

Fig.~2 shows the full-wave scattering efficiency of a plasmonic
cylinder coated with dielectric anisotropic shell versus the
core-shell ratio $a/b$ at different sizes. First, for small coated
cylinder, the long wavelength limit is valid, and hence one can
resort to effective medium theory. In this connection, we discuss
the effective permittivity (effective permeability is unity
because of nonmagnetic components) in Fig.~2(a). We find that for
certain radius ratios $a/b$, the effective permittivity equals $\varepsilon_0$
and the scattering cross section of the coated cylinder is almost
zero correspondingly (see the dips in Fig.~2(b)). As a
consequence, the coated cylinder is invisible or transparent. On
the contrary, when the effective permittivity equals $-\varepsilon_0$ in
Fig.~2(a), the surface plasmon resonance takes place and strong
scattering cross section is obtained (see the corresponding peaks
in Fig.~2(b)). Then, we calculate the scattering efficiency for
larger objects, which are shown in Fig.~2(c) and 2(d). In this
situation, the long wavelength limit is not valid, and effective
medium model cannot be used. In fact, with the object's size being increased, the scattering efficiency can still be reduced significantly (see the dotted lines in Fig.~2(c) and 2(d)) at the certain radius ratio (we call
``near-zero scattering" ratio), though these values are larger than those within the long wavelength limit.

\begin{figure}[htbp]
\begin{center}
\includegraphics[width=12cm]{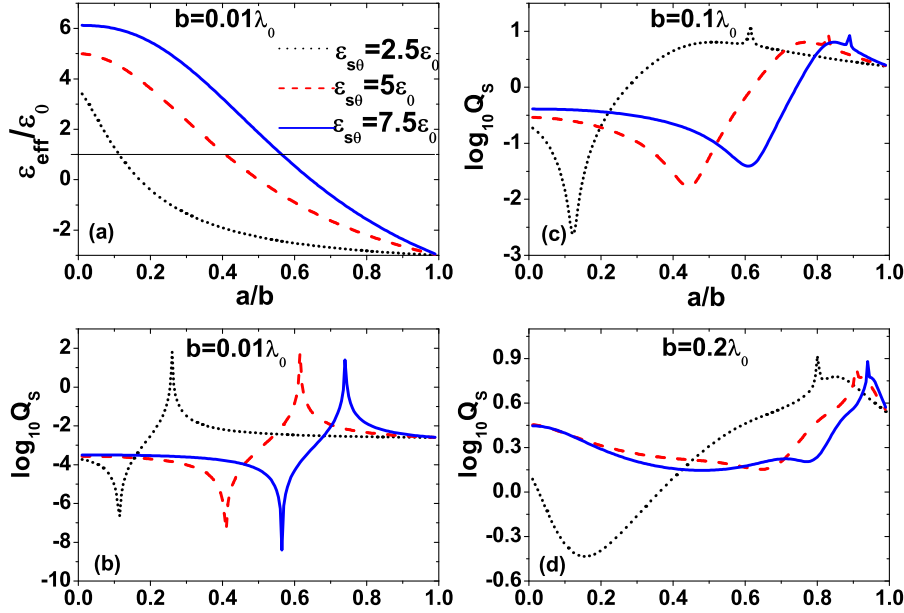} \caption{(a)The effective
permittivity $\varepsilon_{eff}$ of the coated cylinder versus
core-shell ratio for different $\varepsilon_{s\theta}$ when
$b=0.01\lambda_0$. (b)-(d) The scattering efficiency versus
core-shell ratio for different $\varepsilon_{s\theta}$ and
different values of b. $ \varepsilon _{s\theta }=2.5\varepsilon
_0$ (black dotted line); $ \varepsilon _{s\theta }=5\varepsilon
_0$ (red dashed line); $ \varepsilon _{s\theta }=7.5\varepsilon
_0$ (blue solid line), Other parameters are $ \varepsilon _{cr} =
\varepsilon_{c\theta } = - 3\varepsilon _0 ,\quad\varepsilon _{sr}
= 5\varepsilon _0 ,\quad\mu _{sz} = \mu _{cz}  = \mu _0
$.}\label{Fig2}
\end{center}
\end{figure}

Of particular interest is that significant
reduction arises if we tune the value of $\varepsilon_{s\theta}$ in the shell. In Fig.~2(c), with decreasing $\varepsilon_{s\theta}$ (i.e., be
reduced to 2.5$\varepsilon_0$), the scattering efficiency can be
reduced considerably, and thus the transparency or
nearly ``invisible" is attained for large objects.
Fig.~2(d) have verified this tendency at an even
larger size. In other words, through adjusting the dielectric
anisotropy of the shell, it is helpful for us to realize much
lower scattering cross section and better electromagnetic
invisibility. Moreover, the near-zero scattering radius ratio can
be tuned at the same time. We note that a resonant peak is also
presented in Fig.~2, in the long wavelength limit (or for small
size), the maximum of scattering is due to the resonance of $ D_{
\pm 1}^{TM} $. In comparison with the small objects, the
additional small peaks in Fig.~2(c) and (d) result from the
contributions of $ D_{ \pm 2}^{TM} $ because the higher terms of
the scattering coefficient cannot be neglected due to the increasing size of the objects.

\begin{figure}[htbp]
\begin{center}
\includegraphics[width=12cm]{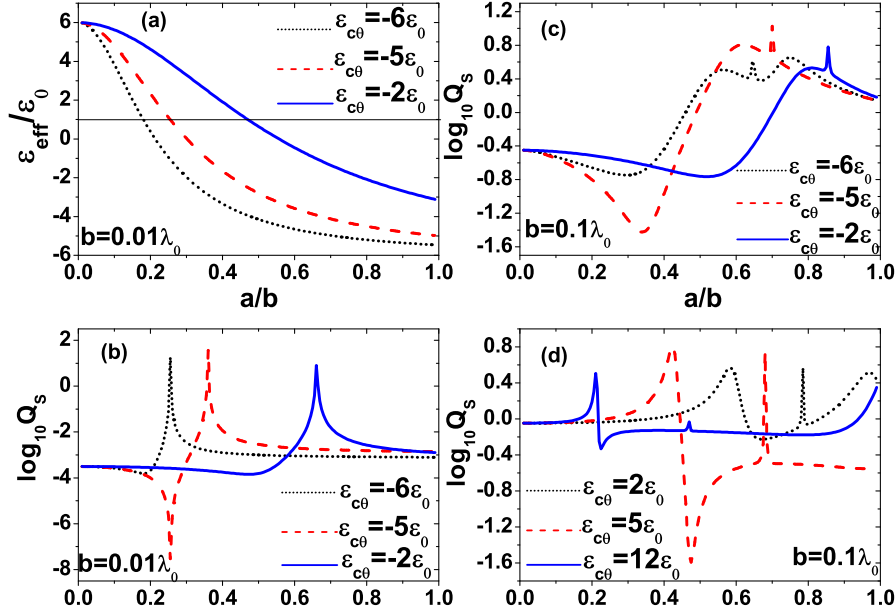} \caption{(a) The effective
permittivity $\varepsilon_{eff}$ of the coated cylinder versus
core-shell ratio for different $\varepsilon_{c\theta}$ when
$b=0.01\lambda_0$. (b)-(c) the scattering efficiency versus
core-shell ratio for different values of b and different
$\varepsilon_{c\theta}$. The values of $\varepsilon_{c\theta}$ in
(a)-(c) are: $ \varepsilon _{c\theta }=-6\varepsilon _0$ (black
dotted line); $ \varepsilon _{c\theta }=-5\varepsilon _0$ (red
dashed line); $ \varepsilon _{c\theta }=-2\varepsilon _0$ (blue
solid line), while other parameters in (a-c) are $\varepsilon
_{cr} = - 5\varepsilon _0 ,\quad\varepsilon _{s\theta }  =
\varepsilon _{sr} = 6\varepsilon _0 ,\quad\mu _{sz} = \mu _{cz} =
\mu _0$. (d) the full-wave scattering efficiency versus core-shell
ratio for $b=0.1\lambda_0$ with $ \varepsilon _{c\theta
}=2\varepsilon _0$ (black dotted line), $ \varepsilon _{c\theta
}=5\varepsilon _0$ (red dashed line), and $ \varepsilon _{c\theta
}=12\varepsilon _0$ (blue solid line), while the other parameters
are $ \varepsilon _{cr} = 5\varepsilon _0 ,\quad\varepsilon
_{s\theta } = \varepsilon _{sr} =  - 10\varepsilon _0 ,\quad\mu
_{sz}  = \mu _{cz}  = \mu _0$.}\label{Fig3}
\end{center}
\end{figure}

Next, we consider a coated cylinder with radially anisotropic core
but isotropic shell as shown in Fig.~3.  Again, for a small coated
cylinder, it is evident that the minimum of the scattering cross
section takes place at the core-shell radius ratio determined by
the condition $\varepsilon_{eff}=\varepsilon_0$.  Hence the
results based on the full-wave theory should be in accordance with
those from effective medium theory. We can observed from
Figs.~3(b-d) that the isotropic core results in the smallest
scattering efficiency, i.e., better electromagnetic invisibility
(e.g., red dashed curves in Figs.~3 representing isotropic cores
in each case). This fact is further proved for larger sizes (see
Fig.~3(c) and Fig.~3(d)), and even for either plasmonic core with
dielectric shell (see Fig.~3(c)) or dielectric core with plasmonic
shell (see Fig.~3(d)). It reveals that in the core-shell system
incorprating radial anisotropy, the isotropy in the core is a
better choice to minimize the scattering width, resulting in
``good" invisibility.

\begin{figure}[htbp]
\begin{center}
\includegraphics[width=8cm]{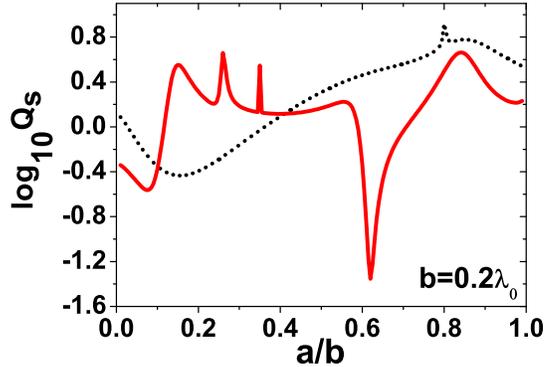} \caption{The full-wave scattering
efficiency versus core-shell ratio for $b=0.2\lambda_0$.
Parameters for the black dotted curve: $ \varepsilon _{c\theta } =
\varepsilon _{cr}  =  - 3\varepsilon _0 ,\quad\varepsilon
_{s\theta }  = 2.5\varepsilon _0 ,\quad\varepsilon _{sr} =
5\varepsilon _0 ,\quad\mu _{sz}  = \mu _{cz} = \mu _0 $.
Parameters for the red solid curve $\varepsilon _{c\theta }  =
\varepsilon _{cr}  =  - 3\varepsilon _0 ,\quad\varepsilon
_{s\theta }  = 2.5\varepsilon _0,\quad\varepsilon _{sr}  =
5\varepsilon _0 ,\quad \mu _{sz}  = 0.5\mu _0 ,\quad\mu _{cz}  = -
7\mu _0 $.}\label{Fig4}
\end{center}
\end{figure}

It is known that for large coated objects, multipolar terms
contribute to the scattering cross section, and the coated
cylinder is visible for almost all the radius ratio (see
Fig.~(4)). In order to make the coated cylinder transparent or at
least less visbile, one can further adjust the magnetic anisotropy
to reduce the scattering cross section.  From Fig.~4, we can
conclude that the invisibility effectiveness of large objects can
be greatly improved through suitable adjustment of both the
dielectric and magnetic anisotropy and the scattering
efficiency can be reduced for nearly one order in magnitude.

\begin{figure}[htbp]
\begin{center}
\includegraphics[width=8cm]{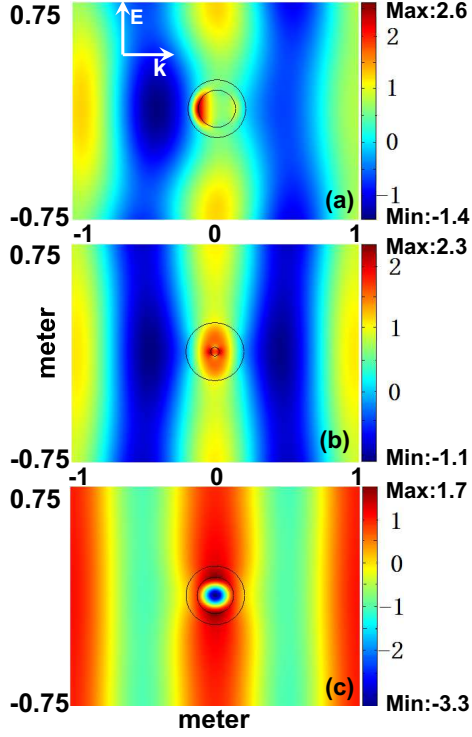} \caption{Snapshots of the real
part of total $H_z$  fields distribution in the $xy$ plane around
the coated cylinder with outer radius $b=0.2\lambda_0$ at 0.3GHz.
(a) Parameters are: $ \varepsilon _{c\theta }  = \varepsilon _{cr}
=  - 3\varepsilon _0 ,\quad\varepsilon _{s\theta } = 5\varepsilon
_0 ,\quad\varepsilon _{sr}  = 5\varepsilon _0 ,\quad\mu _{sz} =
\mu _{cz}  = \mu _0 $, $a=0.645b$. (b)Parameters are: $
\varepsilon _{c\theta }  = \varepsilon _{cr}  =  - 3\varepsilon _0
,\quad\varepsilon _{s\theta } = 2.5\varepsilon _0
,\quad\varepsilon _{sr}  = 5\varepsilon _0 ,\quad\mu _{sz} = \mu
_{cz}  = \mu _0 $, $a=0.155b$. (c) Parameters are: $\varepsilon
_{c\theta }  = \varepsilon _{cr}  =  - 3\varepsilon _0
,\quad\varepsilon _{s\theta } = 2.5\varepsilon _0,\quad
\varepsilon _{sr}  = 5\varepsilon _0 ,\quad\mu _{sz} = 0.5\mu _0
,\quad\mu _{cz}  =  - 7\mu _0 $, $a=0.62b$.}\label{Fig5}
\end{center}
\end{figure}

If the size is sufficiently small, the core-shell ratio determined
by the cloaking condition definitely results in better
invisibility performance. In order to demonstrate the invisibility
performance for a large size $b=0.2\lambda_0$ improved by the
adjustment of the magnetodielectric anisotropy, we present the
patterns of magnetic field in the $xy$ plane, calculated by
finite-element solver of the Comsol Multiphysics. In Fig.~5(a),
for the isotropic case (the same parameters as the red dashed line in Fig.~2(d)), there exists noticeable perturbation caused
by the scattered field. Comparing Fig.~5(b) with 5(a), one can see
that the perturbation is reduced through adjusting the anisotropic
permittivity of the shell. If we take the permeability into
account and properly tune its value, the near ``invisibility"
could be achieved for even relatively larger objects. In
Fig.~5(c), the incident wave is almost unaltered outside the
shelled cylinder, as if there was no scatterer. The
specific core-shell ratios chosen in Fig.~5 are based on the given material parameters, and obtained in numerical calculation for the lowest scattering efficiency correspondingly instead of theoretical effective medium theory (due to the size constraint). It
should be mentioned that the transparency mechanism in this paper
is different from the ideal cloak proposed by Pendry \cite{Pendry}
via spatical compression, where there is no EM field in the core
cylinder. Here the coated cylinder is penetrable.

\section{Conclusion}
In summary, we have demonstrated the full-wave scattering theory
by magnetodielectric anisotropic coated cylinder of infinite
length normally illuminated by a TM polarized plane wave. The
effective permittivity and permeability of the anisotropic coated
cylinder have also been derived in long wavelength limit. The
non-scattering radius ratio obtained from the full-wave theory for
small objects can be well described within the effective medium
theory. Numerical results have shown that the effective scattering
width $ c_{sca}  = 2bQ_{sca}$ can be significantly reduced by
adjusting the anisotropic permittivity of the shell, so as to
achieve better transparency. However it is found that tuning the
core anisotropy is not able to improve the efficiency of
transparency. Furthermore, we could further improve the
invisibility performance of the cylindrical objects by the
adjustment of both electric and magnetic anisotropies. Although we
have only considered the 2D case and losses materials, our
research may be useful for the design of low-observability
cylindrical targets. Based on the above investigation, we may
extend our investigation to more general situations, such as
obliquely incidence with arbitrary polarization, and finite-long
cylinders. We could scale well towards the research of making
multiple cylindrical objects ``invisible".

\section*{Acknowledgments}
This work was supported by  the National Natural Science
Foundation of China under Grant No.~10674098, the National Basic
Research Program under Grant No.~2004CB719801, the Key Project in
Science and Technology Innovation Cultivation Program of Soochow
University, and the Natural Science of Jiangsu Province under
Grant No.~BK2007046.

\newcommand{\noopsort}[1]{} \newcommand{\printfirst}[2]{#1}
  \newcommand{\singleletter}[1]{#1} \newcommand{\switchargs}[2]{#2#1}






\end{document}